\documentclass[aip,rsi,
               amsmath,
               amssymb,
               citeautoscript,
               reprint]{revtex4-2}

\usepackage[colorlinks=true,linkcolor=blue,citecolor=blue]{hyperref}
\RequirePackage[T1]{fontenc}
\RequirePackage[utf8]{inputenc}
\RequirePackage{lmodern}
\usepackage{float}
\usepackage{xfrac}
\usepackage{siunitx}
\usepackage{array}
\usepackage[english]{babel}
\usepackage{textgreek}
\usepackage{xcolor}
\usepackage{braket}

\newcommand{\comment}[1]{}

\def \truevar {1}
\def \falsevar {0}
\def \markupvar {\truevar}

\if\markupvar\falsevar
\newcommand{\changed}[1]{\textcolor{red}{#1}}
\else
\newcommand{\changed}[1]{#1}
\fi

\begin{document}
\selectlanguage{english}

\title{An ultra-stable 1.5 tesla permanent magnet assembly for qubit experiments at cryogenic temperatures}

\author{C. Adambukulam}
\email[Corresponding Author: ]{c.adambukulam@unsw.edu.au}
\affiliation{Centre for Quantum Computation \& Communication Technology, School of Electrical Engineering \& Telecommunications, UNSW Sydney, New South Wales 2052, Australia}

\author{V. K. Sewani}
\affiliation{Centre for Quantum Computation \& Communication Technology, School of Electrical Engineering \& Telecommunications, UNSW Sydney, New South Wales 2052, Australia}

\author{H. G. Stemp}
\affiliation{Centre for Quantum Computation \& Communication Technology, School of Electrical Engineering \& Telecommunications, UNSW Sydney, New South Wales 2052, Australia}

\author{S. Asaad}
\thanks{Currently at Niels Bohr Institute, University of Copenhagen, Blegdamsvej 172100 Copenhagen, Denmark.}
\affiliation{Centre for Quantum Computation \& Communication Technology, School of Electrical Engineering \& Telecommunications, UNSW Sydney, New South Wales 2052, Australia}

\author{M. T. M\k{a}dzik}
\thanks{Currently at QuTech, Delft University of Technology, 2826 CJ Delft, The Netherlands.}
\affiliation{Centre for Quantum Computation \& Communication Technology, School of Electrical Engineering \& Telecommunications, UNSW Sydney, New South Wales 2052, Australia}

\author{A. Morello}
\affiliation{Centre for Quantum Computation \& Communication Technology, School of Electrical Engineering \& Telecommunications, UNSW Sydney, New South Wales 2052, Australia}

\author{A. Laucht}
\email[]{a.laucht@unsw.edu.au}
\affiliation{Centre for Quantum Computation \& Communication Technology, School of Electrical Engineering \& Telecommunications, UNSW Sydney, New South Wales 2052, Australia}

\date{\today}

\begin{abstract}
Magnetic fields are a standard tool in the toolbox of every physicist, and are required for the characterization of materials, as well as the polarization of spins in nuclear magnetic resonance or electron paramagnetic resonance experiments. Quite often a static magnetic field of sufficiently large, but fixed magnitude is suitable for these tasks. Here we present a permanent magnet assembly that can achieve magnetic field strengths of up to \SI{1.5}{T} over an air gap length of \SI{7}{\milli\meter}. The assembly is based on a Halbach array of neodymium (NdFeB) magnets, with the inclusion of the soft magnetic material Supermendur to boost the magnetic field strength inside the air gap. We present the design, simulation and characterization of the permanent magnet assembly, measuring an outstanding magnetic field stability with a drift rate, \changed{$|D|<2.8$~ppb/h}. Our measurements demonstrate that this assembly can be used for spin qubit experiments inside a dilution refrigerator, successfully replacing the more expensive and bulky superconducting solenoids.
\end{abstract}

\pacs{07.55.Db, 75.50.Ww, 03.67.Lx, 76.30.-v, 76.60.-k}


\maketitle 

\section{Introduction}
Experiments that require strong magnetic fields usually rely on superconducting solenoids, which are large, expensive, and require both a stabilized current source and cryogenic temperatures for operation. However, many experiments only need a static magnetic field of order of magnitude \SI{1}{\tesla}, that is ``\textit{set-and-forget}'', as is the case for spin-based quantum computation with electron spin states. With a $g$-factor of $g\approx2$ and a gyromagnetic ratio of $\gamma_e\approx \SI[per-mode=repeated-symbol]{28}{\giga\hertz \per \tesla}$, electron spins in gate-defined quantum dots or donors in silicon~\cite{Ladd2018,Morello2020,Chatterjee2021} require only moderate magnetic fields of $B=0.5-1.5$~T to achieve a Zeeman splitting $\gamma_e B$ that is larger than the thermal energy $k_{\rm B} T$ at typical dilution refrigerator temperatures of $T<100$~mK. This then allows for the read-out and initialization of the spin state via spin-dependent tunneling to and from a thermally-broadened electron reservoir~\cite{Elzerman2004,Morello2010} or via a spin relaxation process~\cite{Watson2018,Yang2020}. Alternative spin initialization and readout methods, like singlet-triplet initialization for quantum dots~\cite{Ono2002,Hanson2007,Yang2020,Hendrickx2020} or optical cycling for colour centers in diamond and silicon carbide~\cite{Doherty2013,Awschalom2018}, do not require $\gamma_e B \gg k_{\rm B} T$, and can therefore work at higher temperatures~\cite{Doherty2013,Yang2020} or with quantum systems that have smaller $g$-factors like electron spins in GaAs quantum dots~\cite{Hanson2007} or hole spins in Ge quantum dots~\cite{Hendrickx2020}. Nevertheless, these systems still require external magnetic fields for selective addressability of individual quantum bits~\cite{Hendrickx2020} or to provide a well-defined quantization axis. In any case, magnetic fields of $B=0.3-1.5$~T are usually sufficient for most quantum computation experiments. 

\newcommand*\circled[1]{{\raisebox{0pt}{\large{\textcircled{\raisebox{0pt} {\small{{#1}}}}}}}}

\begin{figure*}[t]
    \centering
    \includegraphics[scale=1, width=\textwidth]{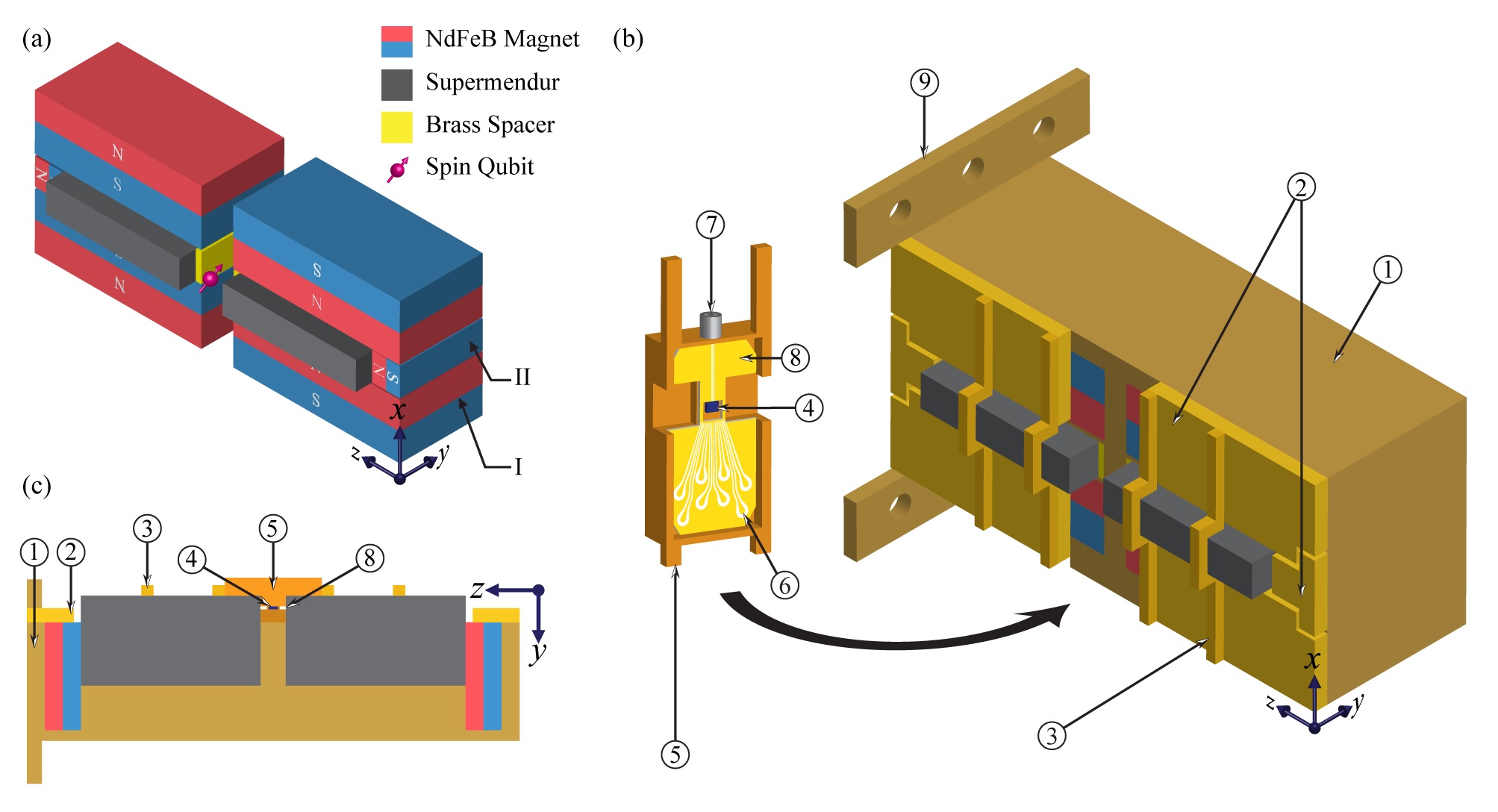}
    \caption{(a) Schematic of the permanent magnet assembly. N and S, and I and II denote north and south, and the two permanent magnet shapes, respectively. (b) The magnet assembly as used in an experiment. The magnets are housed in a copper enclosure \circled{1} with the enclosure lids \circled{2} and brackets \circled{3} used to keep the magnets and Supermendur aligned and in place. The device \circled{4} is housed in its own copper enclosure \circled{5} with DC and MW electrical access provided by DC lines \circled{6} (interfaced with MMCX connectors on the back of \circled{5}) and a coplanar waveguide (interfaced with a 2.92 mm K connector \circled{7}) on a printed circuit board \circled{8}. Thermal contact, in a dilution refrigerator, is ensured by bolting \circled{1} to the mixing chamber plate with the base plate \circled{9}.  (c) Cross-section of the magnet assembly in the $yz$-plane, at the position of the device \circled{4}.}
    \label{fig:setup}
\end{figure*}

In this article we present a permanent magnet assembly that can be used for experiments that require magnetic fields between $0.35$~T and $1.5$~T for samples with footprints of several square millimeters. To achieve these high fields, we employ a design based on Halbach arrays with strong neodymium (NdFeB) permanent magnets, and use the soft magnetic material Supermendur to boost the magnetic field strength even further. The different components are held in place by a copper box with individual compartments and brass lids. One such permanent magnet assembly can be constructed for around $\$750$~USD, made up of $\$100$~USD for the neodymium magnets, $\$150$~USD for the Supermendur pieces and $\$500$~USD for the machined copper box and brass lids.

In Sec.~\ref{sec:design} we show detailed sketches of the design, and use simulations in Sec.~\ref{sec:simulation} to verify that fields of up to $1.5$~T magnitude can be achieved. Furthermore, the fields are tunable by varying the length of the Supermendur pieces and, therefore, the length of the air gap. We then characterise the assembly as a whole, by first presenting cryogenic measurements in Sec.~\ref{sec:cryo} to show that the spin reorientation transition (SRT) of neodymium does not cause any issues, and comparing the cool-down times of a cryogen-free BlueFors dilution refrigerator with a permanent magnet assembly to that of a superconducting magnet. Next, using a single donor spin qubit in silicon - in Sec.~\ref{sec:spin} - we demonstrate that spin qubit experiments can indeed be carried out in the permanent magnet assembly. Monitoring the resonance frequency of the spin qubit allows us to show the long-term stability of the magnetic field \changed{is limited by magnetic field drift, $|D| < 2.8$~ppb/h at the base temperature of our dilution refrigerator. Moreover, this drift} is at least an order of magnitude better than that of \changed{a commercially available superconducting magnet fitted with a state-of-the-art low drift rate option.}

Our proposed design can, therefore, replace traditional superconducting magnets at a significantly lower cost, in experiments where moderate magnetic field gradients can be tolerated. This is for example, the case for Si/Ge quantum dot spin qubits which are routinely controlled via electrically-driven spin resonance inside magnetic field gradients of $\sim 1000$~T/mm generated by micro-magnets \citep{Struck2020}. Moreover, we outline a method to nullify the magnetic field gradients, greatly expanding the applicability of our proposed solution. Additionally, dilution refrigerators can typically only accommodate a single superconducting magnet. Due to the proposed design's light weight, relatively small size, and low stray fields, several magnet assemblies can be mounted in the same fridge and can be used alongside a superconducting magnet. This allows simultaneous measurement of several devices and reduces cool-down times. Further advantages are the stable magnetic field with only a few percent shift in field between room temperature and mK, and the inherent absence of quenching when cryo-cooling fails.

\section{Magnet Assembly}

\begin{figure*}[t]
    \centering
    \includegraphics[scale=1, width=\textwidth]{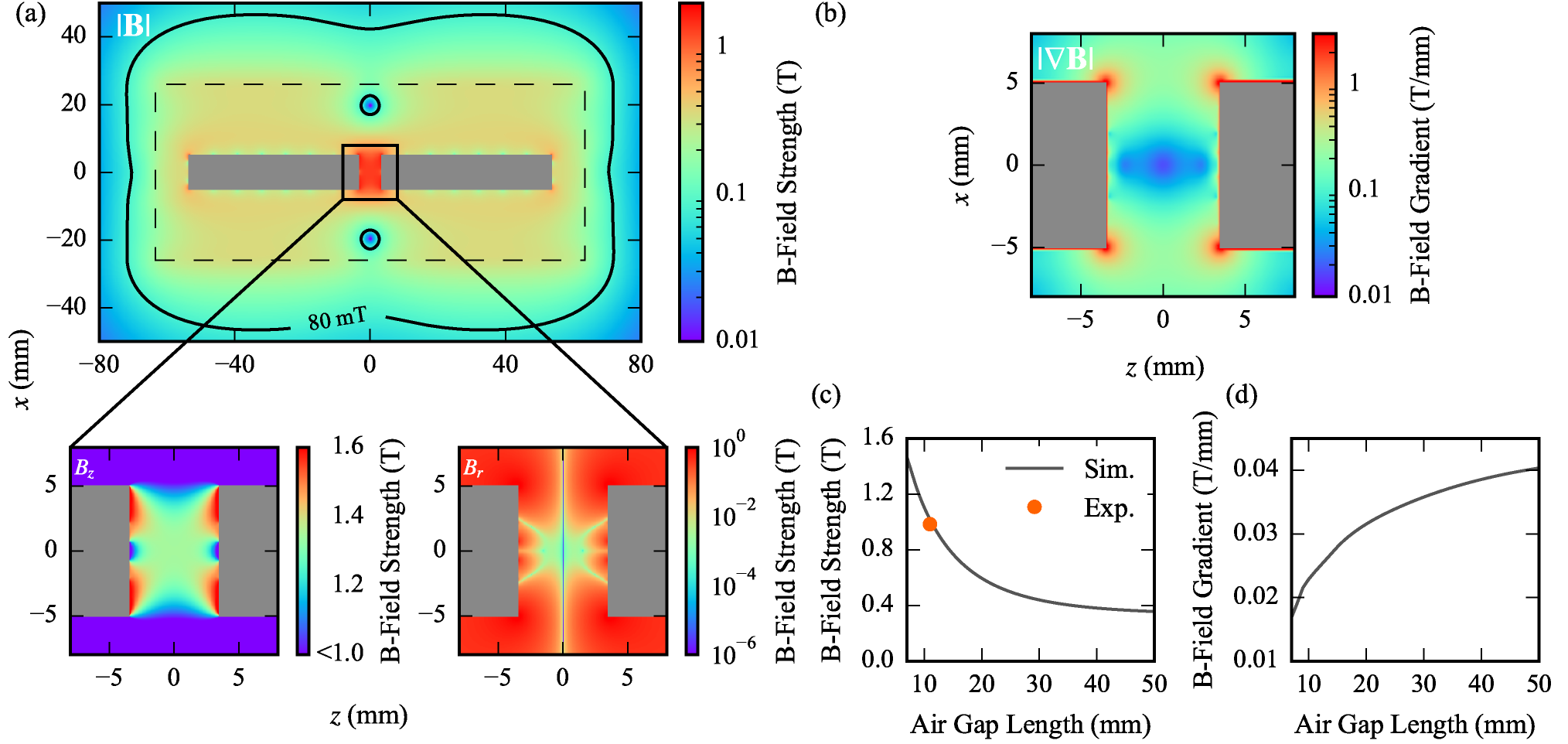}
    \caption{(a) Magnitude of the total magnetic field strength $|\mathbf{B}|$ along an $xz$-cross-section of the magnet assembly, with the location of the qubit taken to be the origin. The dashed line shows the outline of the assembly, the \SI{80}{\milli\tesla} contour marks the critical field of lead and the grey regions denote supermendur. (inset) Magnitudes of \changed{$B_z$} (left) and \changed{$B_r = \sqrt{B_x^2+B_y^2}$} (right) magnetic field components in the air gap. \changed{The desired spin quantization axis is the $z$-axis}. (b) The norm of the magnetic field gradient $|\mathbf{\nabla B}|$. (c) The magnetic field strength at the origin as function of air gap length. The experimental data point is from the measurement shown in FIG.~\ref{fig:spin1}(a). (d) The norm of the magnetic field gradient as a function of air gap length.}
    \label{fig:sim}
\end{figure*}

We base our magnet assembly (see FIG. \ref{fig:setup}) on the design proposed in Ref.~\citenum{Lesnak2012}, a variation of the linear Halbach array~\citep{Halbach1980,Halbach1981,Han2005}. The major point of difference between the design in Ref.~\citenum{Lesnak2012} and the assembly described in this article is the partial replacement of the center magnets with the soft magnetic material, Supermendur (as discussed below) which are then raised out of the magnet plane to provide simple electrical access. A single linear Halbach array is an arrangement of magnets constructed such that the magnetic field is cancelled on one side of the array and enhanced on the other \citep{Choi2008}. In the referenced design, as in our design, two Halbach arrays are placed back-to-back. In this configuration, the magnetic field outside of the array is cancelled and a large and confined magnetic field is generated between the two arrays, and specifically between the two center magnets of the assembly. In the design proposed in Ref.~\citenum{Lesnak2012}, this magnetic field would be at the center of the air gap, which is where devices would need to be mounted. This is a relatively confined space, where it would be difficult to easily fit and route the necessary electrical and microwave connections to operate our devices. This poses difficulties when integrating the assembly into our experimental setup, as to mitigate this issue would require extending the air gap length until the field was impracticably weak or the assembly impracticably large. However, when the Supermendur pieces are raised, a large magnetic field is produced between them at the surface of the assembly, where devices can be placed, with enough space to fit necessary electrical connections. 

In addition, a level of control over the magnetic field strength at the air gap can be attained by the use of brass spacers. The spacers, in conjunction with smaller Supermendur pieces, allow the air gap length and hence, the magnetic field produced by the assembly itself, to be selected at its time of construction. In a more recent, however untested, version of the magnet assembly, we use Supermendur pieces with threaded holes to realize a continuously tunable air gap, that can be tuned by the turn of a screw.

\label{sec:design}

\begin{table}[b]
    \centering
    \begin{tabular}{
        m{2.7cm} m{2.7cm} m{2.7cm}
    }
        \hline \hline
        Magnet & Size (mm) & Elevation (mm)\\
        \hline \hline
        I & $20 \times 30 \times 60$ & 0 \\
        II & $10 \times 30 \times 10$ & 0\\
        Supermendur & $10 \times 25 \times 50$ & 13.5
    \end{tabular}
    \caption{The magnet dimensions (formatted as $x \times y \times z$) and elevation from the bottom of the array (in the negative $y$-direction).}
    \label{tab::dims}
\end{table}

Judicious selection of the magnetic material used is essential in ensuring that the designed assembly produces a sufficiently large magnetic field while minimising its dimensions. As NdFeB magnets are the strongest class of available permanent magnets \cite{Pathak2015,Miyake2018}, N52 grade NdFeB magnets with a typical residual magnetization of approximately \SI{1.45}{\tesla} are used in the assembly. However, at temperatures below \SI{135}{\kelvin}, \changed{it becomes energetically favorable for the magnetic moments in the easy axis of the Nd\textsubscript{2}Fe\textsubscript{14}B phase of the magnet to tilt from its room temperature axis of $[001]$. At $4.2$~K, the easy axis cants $\pm 30.6^\circ$ towards the $[110]$ axis \cite{Tokuhara1985, Givord1984}. This is the well documented spin reorientation transition (SRT) which has the effect of reducing the magnetization of the NdFeB permanent magnet. Consequently, the SRT lowers the magnetic field produced by NdFeB magnet assemblies at cryogenic temperatures \cite{Hara2004,Fredricksen2003}. Nonetheless, it is expected that the magnetization of N52 NdFeB magnets remains large even after the SRT. It should be noted that, following the SRT, although the magnitude of the magnetization will decrease, we expect that due to effect of averaging the tilt over the large number of Nd\textsubscript{2}Fe\textsubscript{14}B grains, the direction of magnetization should remain the same regardless of temperature (see Appendix).}


The inclusion of a soft magnetic material with a low residual magnetization and a high saturation magnetization can be used to further strengthen the magnetic field produced by the assembly. As mentioned above, the center permanent magnets are partially replaced  with Supermendur, a soft magnetic alloy of $2\%$ vanadium, $49\%$ iron and $49\%$ cobalt that has a saturation magnetization of approximately \SI{2.4}{\tesla} \cite{Gould1957}. That is $\sim 1$~T greater than the residual magnetization of NdFeB. Hence, the magnetic field produced by the assembly can be enhanced by placing saturated supermendur near the air gap. To that end, the center magnets are partially replaced by Supermendur which, given its low saturation induction\cite{Gould1957}, is saturated near the air gap. Thus, the Supermendur provides a major contribution to, and hence substantially increases, the magnetic field in the air gap. In addition, based on the negligible decrease in saturation magnetization observed for Supermendur at \SI{4.2}{\kelvin} \cite{Ackermann1971}, we assume that the cryogenic and room temperature behavior of Supermendur are similar. 

As a consequence of the highly confined magnetic field in conjunction with the use of strong magnetic materials, the magnet assembly is compact with a footprint of \SI{55}{} $\times$ \SI{87}{\milli \meter}. For reference, the diameter of the mixing chamber plates of Bluefors LD and XLD series dilution refrigerators are \SI{290}{\milli\meter} and \SI{500}{\milli\meter}, respectively. Thus, we estimate that at least four magnet assemblies can be mounted vertically to one side of the mixing chamber plate  of an LD series dilution refrigerator [that is, mounted with the $z$-axis perpendicular to the mixing chamber plate surface with the point of contact being the base plate of the assembly - \circled{9} in FIG. \ref{fig:setup}(b)]. If both, the top and the bottom surfaces of the mixing chamber plate are fitted with magnet assemblies, that number rises to eight and would further increase for an XLD series dilution refrigerator. This enables multiple experiments to be run in the same dilution refrigerator, with the upper bound on the number of experiments set by the cabling and necessary electronics and not by the bore of a superconducting solenoid. 

\section{Simulation}
\label{sec:simulation}

The magnet assembly was simulated in Radia, a boundary integral method magnetostatics solver for Wolfram Mathematica, developed by the European Synchrotron Radiation Facility \cite{Chubar1998,Elleaume1998}. Given the large coercivity of NdFeB magnets at cryogenic temperatures \cite{Hara2004}, we model the permanent magnets as insensitive to any external magnetic field applied to the magnet and with a residual magnetization of \SI{1.24}{\tesla}. This is the room temperature residual magnetization rotated by $30.6^\circ$ and then projected onto the room temperature easy axis. While this model is not an entirely accurate representation of an NdFeB magnet, especially given the rising magnitude of the residual magnetization below the SRT \cite{Tokuhara1985}, it suffices to estimate the worst case behaviour of the magnets. We model the Supermendur as a non-linear magnetic material based on its room temperature magnetization curve. Finally, to simulate the effect of varying the air gap length, we correct our estimate of the magnetization of the NdFeB magnets at millikelvin temperatures from the electron spin resonance frequency in a silicon-based single-atom spin qubit device~\citep{Muhonen2014, Madzik_thesis, Madzik2021} mounted inside the permanent magnet assembly (see Sec.~\ref{sec:spin}).

\begin{figure}[t]
    \centering
    \includegraphics[scale=1]{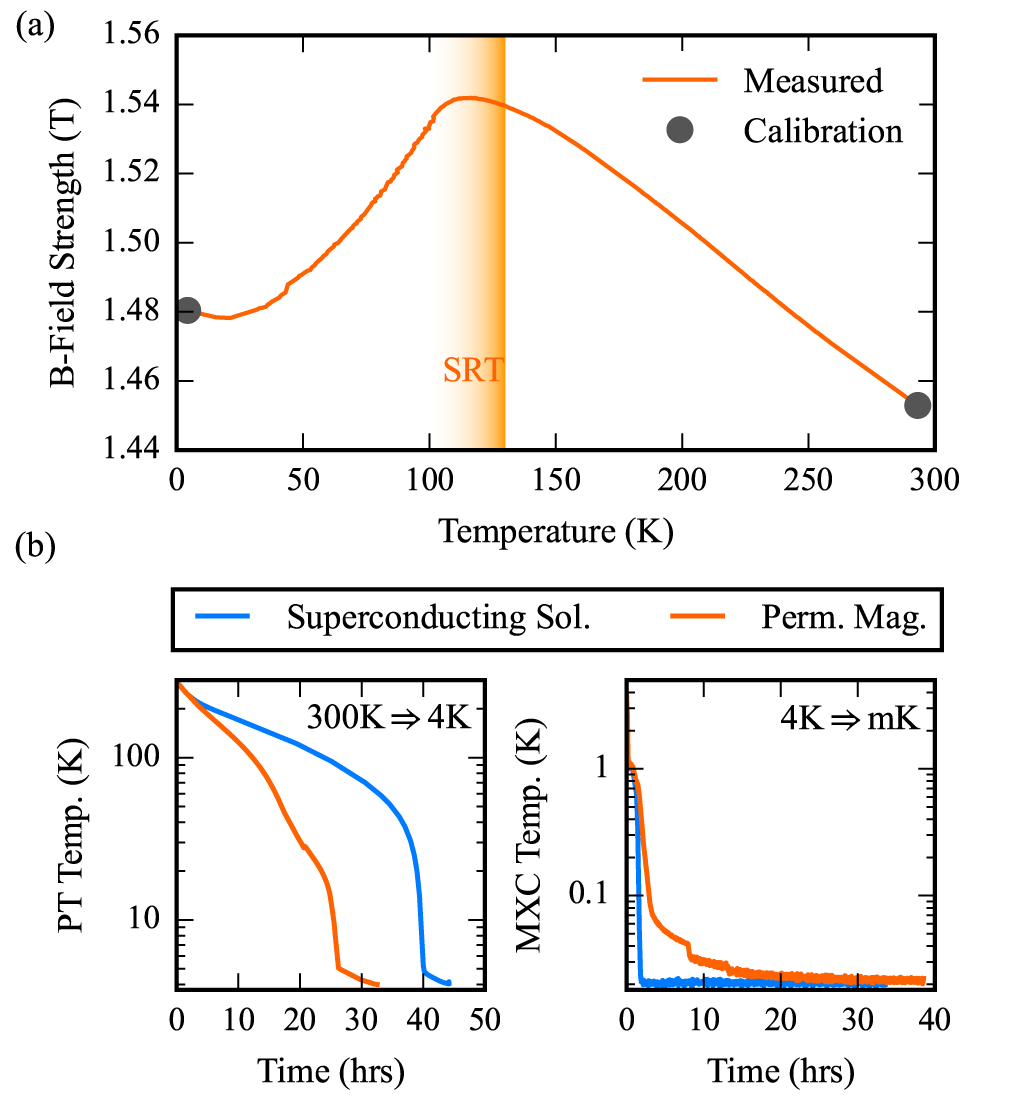}
    \caption{(a) Temperature dependence of the magnetic field generated by the magnet assembly, with the NdFeB spin reorientation transition (SRT) indicated. (b) Pulse tube (PT, left) and mixing chamber (MXC, right) temperature during the two cool-down stages of a dilution refrigerator equipped with either a superconducting solenoid or a permanent magnet assembly.}
    \label{fig:cryo}
\end{figure}

In FIG.~\ref{fig:sim} we plot the results of our magnetic field simulations for an air gap of \SI{7}{\milli\meter} (the assembly is without brass spacers). The main panel in FIG.~\ref{fig:sim}(a) shows $|\mathbf{B}|$ the magnitude of the magnetic field strength for the whole magnet assembly, plotted for the $xz$-cross-section at the $y$-position of the qubit location. The bottom insets are zoom-ins to the qubit region, showing the \changed{$B_z$ and $B_r=\sqrt{B_x^2+B_y^2}$} components with respect to the qubit quantization axis. At the expected qubit position, we compute a magnetic flux density of \SI{1.44}{\tesla} that is strongly aligned to the $z$-axis and strongly confined to the air gap. The high magnetic flux density ensures that the Zeeman splitting of the electron spin is larger than the thermal energy at dilution refrigerator temperatures, a prerequisite for many spin initialization and readout schemes~\cite{Elzerman2004,Morello2010,Watson2018,Yang2020}. In addition, the strong magnetic field confinement to the inside of the enclosure would, in the future, allow for total electromagnetic isolation of the qubit by enclosing the assembly and device inside a superconducting shield. In FIG.~\ref{fig:sim}(a), we mark the \SI{80}{\milli\tesla} iso-field line indicating where the stray field falls below the critical field of lead, to approximate the dimensions required by such a shielding enclosure. The outline of the magnet assembly is indicated by the dashed line. 

In FIG.~\ref{fig:sim}(b) we plot the \changed{norm} of \changed{the} magnetic field gradient \changed{which we compute as $|\mathbf{\nabla B}| = \sqrt{\sum_{i,j} (\partial_i B_j)^2}$ where $i,j \in \{x,y,z\}$}. We \changed{find $|\mathbf{\nabla B}| \sim 18$~mT/mm} at the qubit location. This is considerably greater than that which can be achieved by shimmed superconducting solenoids $< 1 $~\textmu T/mm \citep{Britton2016} or even $\sim 5$~\textmu T/mm - the value simulated for the superconducting solenoid used in our setup when generating a $1.55$~T magnetic field \cite{Kalra2016}.  A more detailed inspection reveals that the only non-zero gradient components are $\partial_y B_z = \partial_z B_y \approx -12.6$~mT/mm. Nullification of \changed{this gradient term} can be achieved by lowering the elevation of the Supermendur pieces from $13.5$~mm to $2.5$~mm, thereby, introducing mirror symmetry in the $xz$-plane. Samples must then be placed, at the centre of the assembly (at an elevation of $15$~mm from the bottom of the array). However, this approach comes with the disadvantage of difficult electrical access. While the \changed{moderate} magnetic field \changed{gradient} would be detrimental to the performance \changed{of spatially extended qubit ensembles and spatially separated qubits}, they are not detrimental to single donor spins in silicon. In our setup a magnetic field gradient can be transduced into magnetic noise on the qubit if the magnet moves with respect to the sample stage~\cite{Britton2016,Kalra2016,Struck2020}. This is to be expected in common cryomagnetic systems, where the sample plate and the magnet are connected at very distant flanges - as is the case with superconducting solenoids. The matter can be particularly severe in cryogen-free refrigerators due to the vibrations produced by the pulse-tube coolers. However, in the case of the permanent magnet assembly, the magnet and sample form part of the same compact and rigid structure. Relative motion between the magnet assembly and the sample is suppressed, making the field gradient inconsequential for qubit coherence. 

Finally, in FIG.~\ref{fig:sim}(c) and FIG.~\ref{fig:sim}(d) we show how the magnetic field strength and gradient at the qubit position vary when the size of the air gap length is increased. This can be accomplished for example by shortening the Supermendur pieces and inserting brass spacers near the air gap to keep them in place. The simulations show that the field strength can be adjusted from \SI{1.44}{\tesla} to \SI{0.36}{\tesla} when the air gap length is varied from $7$~mm to $50$~mm. The upper limit to the produced magnetic field is $\sim 2.74$~T and is the sum of the saturation magnetization of Supermendur and the magnetic field provided by the rest of the assembly. This upper limit is achieved when the air gap length approaches $0$~mm. In order to ensure a large enough sample volume to comfortably mount devices, we limit the minimum air gap length to $7$~mm. The air gap length can be extended - by shortening the Supermendur pieces - until no Supermendur remains. At this point the assembly produces $\sim 0.35$~T. This provides a large range of magnetic fields for qubit experiments, albeit without the possibility for in-situ tunability of the assembly in its current form.

\section{Experiments}
\label{sec:exp}

\subsection{Cryogenic Behaviour of the Assembly}
\label{sec:cryo}

\begin{figure}[t]
    \centering
    \includegraphics[scale=1]{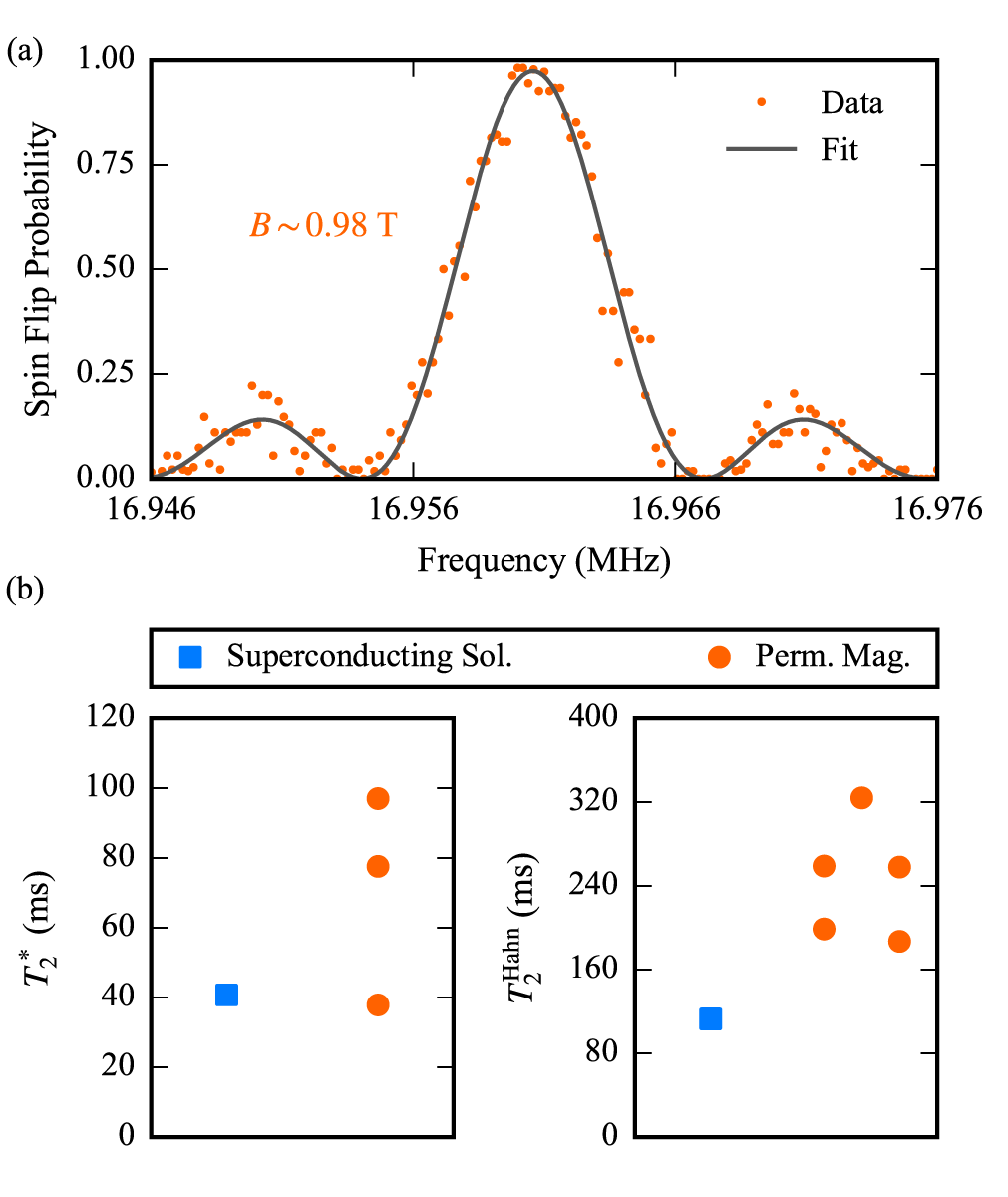}
    \caption{
    (a) Coherent nuclear magnetic resonance (NMR) spectrum of the ionized \textsuperscript{31}P$^+$ donor nuclear spin performed with a permanent magnet assembly. 
    (b) Free induction decay $T_2^*$ (left) and spin echo $T_2^{\rm Hahn}$ (right) times for ionized \textsuperscript{31}P$^+$ donor nuclear spins from various devices of the latest batches~\cite{Madzik_thesis}. The measurements were performed with either a permanent magnet assembly or a superconducting electromagnet and were performed on different devices leading to the spread in the measured $T_2^*$ and $T_2^{\rm Hahn}$ values\changed{.}}
    \label{fig:spin1}
\end{figure}

To measure the temperature response of the magnet assembly, a Cryomagnetics Inc. HSU-1 calibrated Hall effect sensor is placed in the air gap and the assembly is installed inside a dilution refrigerator. The Hall voltage is measured during warm up of the fridge. FIG.~\ref{fig:cryo}(a) shows the temperature response of the assembly with the NdFeB spin reorientation transition (SRT) visible. We attribute the relatively small change in magnetic field over the temperature range to the saturated Supermendur, which produces the bulk of the magnetic field at the air gap and which is insensitive to fluctuations in the NdFeB magnetization. Overall, the magnetic field inside the air gap changes by only $\sim6\%$ over the whole temperature range from \SI{300}{K} down to \SI{}{mK},\footnote{It should be noted that the measured dependence in FIG.~\ref{fig:cryo}(a) is produced by assuming a first order temperature response of the Hall sensor between the two calibrated points. Moreover, the Hall sensor, while placed in the air gap, could not be placed at the expected qubit location. Nonetheless, this measurement demonstrates the relative changes in magnetic field strength and given the high homogeneity of the field in the air gap, the measured field should be similar to the field experienced by a spin qubit.} making it an interesting option for wide-range temperature dependent studies and measurements.

Next, we measure the cool-down time of a BlueFors BF-LD400 dilution refrigerator with either a permanent magnet assembly or an American Magnetics 5-1-1~T superconducting vector magnet mounted. We observe a $\sim$ \SI{14}{\hour} improvement in the time taken for the pre-cool step from room temperature to \SI{4}{\kelvin} [see FIG.~\ref{fig:cryo}(b) - left panel]. This is expected as the mass of the superconducting solenoid (\SI{38.1}{\kilo \gram}) is considerably larger than that of the assembly (\SI{2.83}{\kilo \gram}). When using the superconducting magnet, the cooldown time from $4$~K to $30$~mK is $\sim 8$~h shorter than that of the permanent magnet assembly. As the superconducting solenoid operates at \SI{4}{\kelvin} it is not further cooled during the second cool-down step from \SI{4}{\kelvin} to \SI{}{\milli\kelvin}, however, the magnet assembly is cooled to \SI{}{\milli\kelvin} along with the sample. This, along with the weaker cooling powers at lower temperatures, accounts for the longer cooling time in the second cool-down step when the magnet assembly is used [see FIG.~\ref{fig:cryo}(b) - right panel]. Nonetheless, there is a $\sim 6$~h improvement in total cool-down time when using the permanent magnet assembly.

There is one further distinct advantage of the permanent magnet assembly compared to a superconducting magnet. Superconducting magnets may quench and raise the cryostat temperature by tens of kelvin. This situation is particularly common in cryogen-free dilution refrigerators that are critically dependent on a chilled water supply. The helium compressor of the pulse tube will shut off within tens of seconds of any interruption to the cooling water resulting in the overheating and quenching of the superconducting solenoid. This is not an issue with the permanent magnet assembly.

\subsection{Spin Qubit Experiments}
\label{sec:spin}

As a final test for the permanent magnet assembly, we use it to provide the magnetic field for spin qubit experiments with ion implanted, single $^{31}$P donors in isotopically enriched $^{28}$Si (see Appendix for more details).  The physical systems and devices have been described in ample detail elsewhere~\cite{Morello2010,Muhonen2014,Madzik2021}, and we simply use them to probe the magnetic field and its stability. 

In FIG.~\ref{fig:spin1}(a) we show a coherent nuclear magnetic resonance (NMR) spectrum of the ionized $^{31}$P$^+$ nuclear spin. In this charge configuration, the nucleus has no hyperfine coupling. The NMR frequency is thus given by $\nu_{\rm{NMR}}=\gamma_n B$, where $\gamma_n = \SI[per-mode=repeated-symbol]{-17.23}{\mega\hertz \per \tesla}$ is the nuclear gyromagnetic ratio of $^{31}$P. By performing repeated, non-destructive single-shot readout while varying the frequency of an RF magnetic field, we \changed{can measure the $z$-component of the applied magnetic field to a high degree of precision}. We extract \changed{$B_z = 0.984082 \pm 0.000004 $~T}, which is in good agreement with our simulations, as shown in FIG.~\ref{fig:sim}(c).

\begin{figure}[H]
    \centering
    \includegraphics{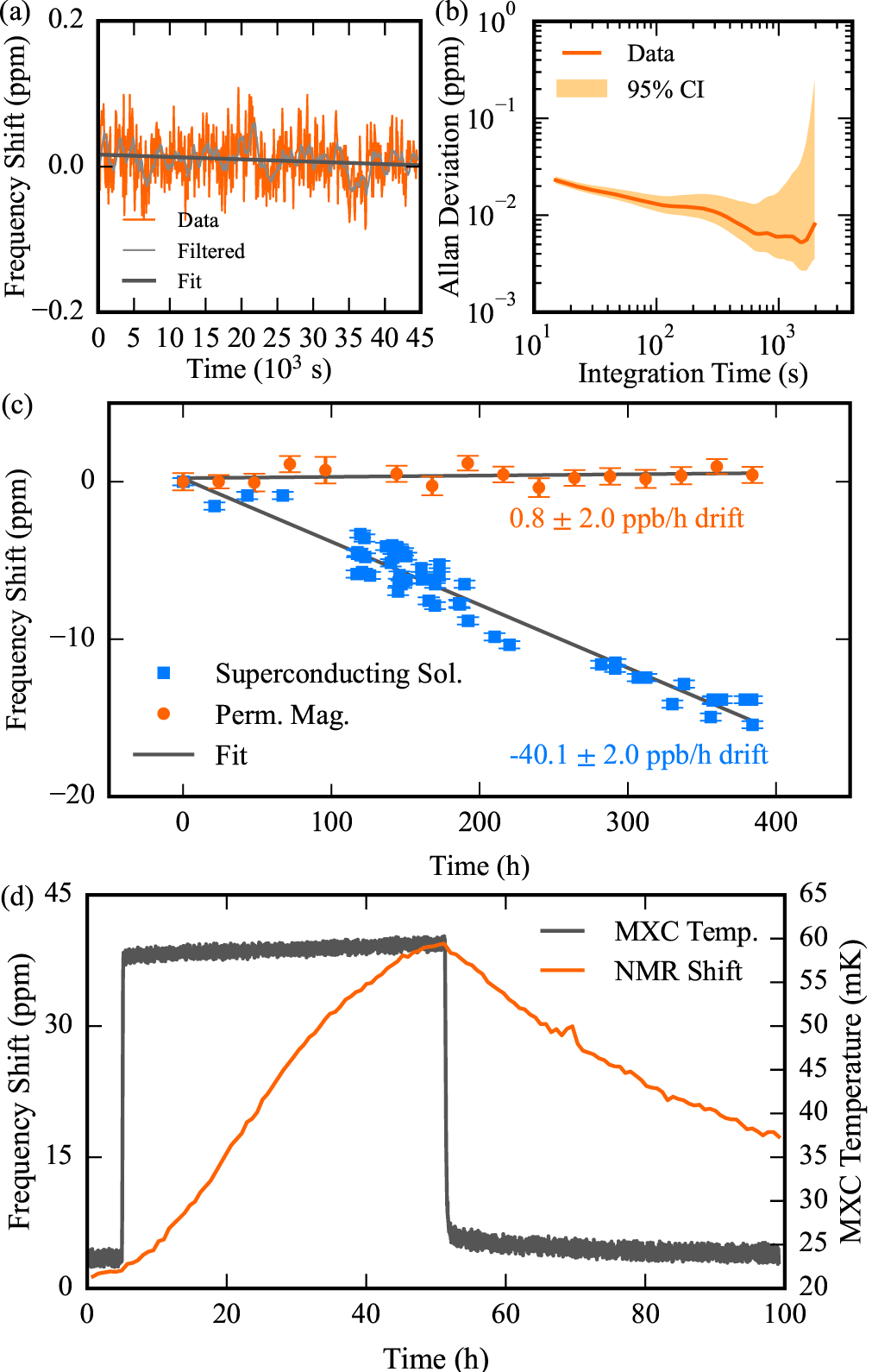}
    \caption{\changed{
    (a) Shift in the NMR frequency of a $^{31}$P$^+$ nuclear spin measured over the course of $\sim 13$~h. The measurement was performed using the permanent magnet assembly. The data is fit to a line and a drift rate of $-1.2 \pm 0.8$~ppb/h is extracted. A $10$-point rolling average filtered version of NMR shift (light grey curve) has been provided to make the drift in the NMR frequency shift more apparent.
    (b) The overlapped Allan deviation and its $95\%$ confidence interval (CI) of the data presented in (a) as function of integration time, $\tau$. 
    (c) The long term shift in the resonance frequency of a $^{31}$P$^+$ nuclear spin and a $^{31}$P$^0$ electron spin in the permanent magnet assembly (orange circles) and a superconducting solenoid (blue squares), respectively.
    (d) The shift in NMR frequency and mixing chamber (MXC) temperature as function of time. After time $t=4$~h, the MXC heater is turned on, resulting in the rise in MXC temperature. At $t=51$~h, the MXC heater is turned off and the temperature returns to the dilution refrigerator base temperature.}}
    \label{fig:spin2}
\end{figure}

We try to gauge possible fluctuations in the magnetic field produced by the assembly by conducting coherence time measurements on the nuclear spin. Unfortunately, we were not able to compare the same device and donor in both a superconducting solenoid and a permanent magnet assembly. Hence, we show in FIG.~\ref{fig:spin1}(b) the free induction decay coherence times $T_2^*$ and Hahn echo coherence times $T_2^{\rm{Hahn}}$ for a few devices belonging to similar, recent fabrication batches~\cite{Madzik_thesis}. While the outcome of this comparison is certainly not conclusive, the data suggests that the permanent magnet assembly does not worsen the qubit coherence, and possibly improves it compared to a superconducting solenoid. \changed{A more conclusive test of the possible fluctuations of the permanent magnet assembly would be to perform spin noise spectroscopy, as performed in} Ref. \citenum{Muhonen2014} \changed{using the electron spin of a single $^{31}$P donor. In that system, however, the frequency fluctuations due to magnetic field instability would be compounded with other effects, caused by, for example, residual coupling to $^{29}$Si nuclei or electric field noise affecting the electron g-factor or electron-nuclear hyperfine coupling \citep{Laucht2015}. For this reason, we have chosen to utilize the $^{31}$P nucleus as our magnetic field probe. However, because of the much longer ($\sim 3$ orders of magnitude) coherence time of the nuclei compared to electrons, nuclear spin noise spectroscopy would take an unmanageable amount of time and has indeed never been attempted.}

\changed{Finally, we attempt to determine the long term stability of the permanent magnet assembly using the NMR frequency of a \textsuperscript{31}P$^+$ nuclear spin. Previously we attempted to gain insight into magnetic field fluctuations that occur in the $1$~Hz to $20$~kHz regime; which strongly affect qubit coherence and therefore may be detected by the free induction decay or spin echo. Here, we measure the magnetic field drift and fluctuations over the course of hours and days. To gauge the magnetic field stability, we compute the overlapped Allan deviation} \citep{Howe2008} \changed{as presented in FIG.~\ref{fig:spin2}(b) from the shift in NMR resonance frequency shown in FIG.~\ref{fig:spin2}(a).  We measure a minimum Allan deviation on the order of $\sim 10^{-9}$ when the integration time $\tau \sim 1500$~s.} 

\changed{To investigate the drift, we fit the NMR frequency trace produced in the FIG.~\ref{fig:spin2}(a) to a line and extract a drift rate, $D = -1.2 \pm 0.8$~ppb/h. Additionally, we measure the shift in NMR frequency over the course of $17$~days [see FIG.~\ref{fig:spin2}(c)] but were unable to resolve any drift beyond the fitting error. Thus, we set an upper bound on the magnetic field drift rate to $|D| < 2.8$~ppb/h.} 

\changed{A possible source of the magnetic field instability is the drift and fluctuations of the temperature of the permanent magnet assembly. To that end we measure the shift in the NMR frequency [see FIG.~\ref{fig:spin2}(d)] when the MXC heater is turned on and the mixing chamber plate temperature rises. We observe a $\sim 40$~ppm shift in the NMR frequency when the MXC temperature reaches $\sim 60$~mK. The significant lag in the NMR shift compared to the change in the MXC temperature suggests that the effect is indeed due to the very slow change in the magnetic field while the permanent magnet assembly reaches thermal equilibrium. In principle, it is conceivable that some mechanisms internal to the device (paramagnetic spins, Pauli magnetism in electron gases) might cause minuscule magnetic fields which are also temperature dependant, but we would expect such effects to react much more quickly to temperature change at the mixing chamber. In any case, the results presented here provide ample evidence that the normal temperature stability of a dilution refrigerator results in remarkably low magnetic field drift rates.} In previous measurements with standard superconducting solenoids in a liquid helium bath and operated in persistent mode, we noticed a decay of $\sim15$~ppm/h when the driving current through the leads was off, and $\sim1.5$~ppm/h when the nominal driving current was still fed through the leads\cite{Muhonen2014}. Using an American Magnetics superconducting magnet with low drift rate option for the persistent mode switch reduces the magnetic field decay to $\sim 40$~ppb/h [see FIG.~\ref{fig:spin2}(c)]. This corresponds to a change in the $40$~GHz resonance frequency of the electron spin qubit of about 1~MHz in a month. The drift rate of the permanent magnet assembly is over an order of magnitude lower than the drift measured with the state-of-the-art superconducting solenoid, and compares favourably to the magnetic field drift specified for commercially available NMR spectrometers of $\sim10$~ppb/h~\cite{Bruker} and the $\sim 1$~ppb/h that are practically possible~\cite{Kemp1986}. \changed{Moreover}, the excellent magnetic field stability is a strong indication that the use of permanent magnet assemblies like the one described here can be a key enabling technology to achieve the best possible long-term stability and for qubits whose energy splitting depends upon a magnetic field. 

\section{Conclusion}
In conclusion, we have presented a permanent magnet assembly based on neodymium (NdFeB) magnets that can provide magnetic field strengths of up to \SI{1.5}{T} over an air gap of \SI{7}{\milli\meter} length. The assembly works for a wide temperature range from \SI{300}{K} to \SI{}{\milli\kelvin} temperature with only a few percent variation in magnetic field strength. This makes it ideal for a wide variety of experiments, including spin qubits, nuclear magnetic resonance, and electron paramagnetic resonance spectroscopy, where medium-high fields with good stability are required. 
Furthermore, with a production price of $\sim\$750$~USD, a size of $138 \times 87 \times 55$~mm, and a weight of $2.83$~kg, the assembly is much cheaper and more compact than superconducting solenoids, and allows several experiments to be run in parallel on the same mixing chamber plate of a dilution refrigerator. 

\begin{acknowledgments}
We acknowledge the team involved in the fabrication of the donor spin qubit devices used for the field stability test: Fay E. Hudson, Kohei M. Itoh, David N. Jamieson, A. Melwin Jakob, Brett C. Johnson, Jeffrey C. McCallum, and Andrew S. Dzurak. The research was funded by the Australian Research Council Centre of Excellence for Quantum Computation and Communication Technology (Grant No. CE170100012) and the US Army Research Office (Contract No. W911NF-17-1-0200). We acknowledge the support of the Australian National Fabrication Facility (ANFF). A.L. and C.A. acknowledge support through the UNSW Scientia Program. The views and conclusions contained in this document are those of the authors and should not be interpreted as representing the official policies, either expressed or implied, of the ARO or the US Government. The US Government is authorized to reproduce and distribute reprints for government purposes notwithstanding any copyright notation herein.
\end{acknowledgments}

\section*{Data Availability Statement}
The data that support the findings of this study are available from the corresponding authors upon reasonable request.
\\

\section*{Appendix}
\addcontentsline{toc}{section}{Appendices}
\renewcommand{\thesubsection}{\Alph{subsection}}

\setcounter{equation}{0}
\renewcommand\theequation{C\arabic{equation}}

\subsection{Cryogenic Magnetization of NdFeB Magnets}

\changed{As a quick test to check whether the magnetization direction of the NdFeB magnets experienced a change in magnetization direction when cooled, a Hall effect sensor was placed at the center of three orthogonal surfaces of a sample magnet and the magnetic field was recorded at room temperature and at $1.5$~K (see Tab.~\ref{tab::N55}). While this measurement cannot be used to determine the cryogenic response of the magnet, it serves to demonstrate that no significant change in magnetization axis occurs.}  

\begin{table}[H]
    \centering
    \begin{tabular}{
       m{2.7cm}  m{3.2cm}  m{2cm}  
    }
        \hline \hline
        \vspace{3pt}
        Surface & B at \SI{300}{\kelvin} (T) & B at \SI{1.5}{\kelvin} (T) \\
        \hline \hline
        $\langle x \rangle$ & 0.00723 & 0.00663\\
        $\langle y \rangle$ & 0.576 & 0.449\\
        $\langle z \rangle$ & 0.0653 & 0.0737\\
    \end{tabular}
        
    \caption{Surface magnetic flux density of a $20 \times 30 \times 60$~mm N55 NdFeB magnet magnetized along the $y$-direction, at \SI{300}{\kelvin} and \SI{1.5}{\kelvin}. While the magnitude reduces by $\approx 20\%$, the magnetization direction does not change. Note, that the NdFeB magnets of this size that are used in the assembly are actually magnetized in the $x$-direction.}
    \label{tab::N55}
\end{table}

\subsection{Device Information}
\label{app::device_information}

The devices used in these experiments consist of ion implanted $^{31}$P donors in isotopically enriched $^{28}$Si with a $^{29}$Si concentration of $\sim 800$~ppm. For measurements with samples mounted to a superconducting magnet, P$_2^+$ molecular ions were implanted such that approximately four donor pairs are created in a $90 \times 90$~nm\textsuperscript{2} window. Measurements performed on samples mounted to a permanent magnet assembly were implanted with P$^+$ ions to yield a donor concentration of $1.25 \times 10^{12}$~donors/cm\textsuperscript{2}. We estimate a mean donor separation of $8$~nm and $4$~nm for the former and latter samples, respectively.

\subsection{Measuring the Magnetic Field with a Single Nuclear Spin}

\changed{The shift in NMR frequency presented in FIG.~\ref{fig:spin2}(a) was measured by \textit{side-of-fringe} spin magnetometry} \citep{maze2008,taylor2008}. \changed{In such a scheme, following the initialization of the spin to $x_i = \ket{\Uparrow}/\ket{\Downarrow}$ by measurement of $S_z$, the spin is put into the state (in the rotating frame and up to a global phase)} 

\begin{equation}
    \ket{\psi} = \frac{1}{\sqrt{2}}\big( \ket{\Downarrow} \pm i\ket{\Uparrow} \big)
\end{equation}

\changed{by a $\frac{\pi}{2}$ rotation about the $x-axis$. The spin is then left to precesses in the $xy$-plane of the Bloch sphere for a fixed time, $\tau$. If there is a detuning, $\Delta = \omega - \omega_0$ where $\omega$ is the NMR frequency and $\omega_0$ is the frequency of the rotating frame, then following the time $\tau$ the spin is in the state}

\begin{equation}
    \ket{\psi} = \frac{1}{\sqrt{2}}\big( e^{i\frac{\Delta\tau}{2}}\ket{\Downarrow} \pm ie^{-i\frac{\Delta\tau}{2}}\ket{\Uparrow} \big).
\end{equation}

\changed{A subsequent $\frac{\pi}{2}$ rotation about the $y-axis$ is used to put the spin - up to a global phase - to either,}

\begin{equation}
    \ket{\psi} = \cos{\bigg(\frac{\Delta\tau}{2}-\frac{\pi}{4}}\bigg) \ket{\Downarrow} -i\sin{\bigg(\frac{\Delta\tau}{2}-\frac{\pi}{4}}\bigg) \ket{\Uparrow},
\end{equation}

\changed{if the initial state $x_i = \ket{\Uparrow}$ or}

\begin{equation}
    \ket{\psi} = i\sin{\bigg(\frac{\Delta\tau}{2}-\frac{\pi}{4}}\bigg) \ket{\Downarrow} -\cos{\bigg(\frac{\Delta\tau}{2}-\frac{\pi}{4}}\bigg) \ket{\Uparrow},
\end{equation}

\changed{if $x_i = \ket{\Downarrow}$. We find that the spin flip probability,  $P_\mathrm{flip}$ - that is the probability of measuring $x_m = \ket{\Uparrow}$ given $x_i = \ket{\Downarrow}$, or vice versa - is given by}

\begin{equation}
    P_\mathrm{flip} = \frac{1}{2}+\frac{1}{2}\sin{\Delta\tau},
\end{equation}

\changed{or equivalently,}

\begin{equation}
    \Delta \approx \frac{2P_\mathrm{flip}-1}{\tau}.
    \label{eq:mag}
\end{equation}

\changed{From Eq.~\ref{eq:mag}, it is apparent that the sensitivity is improved with increasing $\tau$. However, in practice we are limited by decoherence which adds a factor of $e^{-\chi(\tau)}$ - where $\chi$ is the coherence function - to the denominator of Eq.~\ref{eq:mag}. As we measure $T_2^* = 51.1 \pm 2.5$~ms and $\chi(\tau) = (\tau/T_2^*)^2$, we set $\tau = 25$~ms as a compromise.}

\changed{It can be shown} \citep{Degen2017} \changed{that the uncertainty in $P_\mathrm{flip}$ is given by}

\begin{equation}
    \frac{1}{2 C \sqrt{N}},
\end{equation}

\changed{where $C$ is dimensionless efficiency parameter, dependent on the spin readout fidelity and $N$ is the number of repetitions performed. For our $^{31}$P$^+$ nuclear spin, $C \sim 1$ and $N = 300$, resulting in an error in $\Delta$ of $0.23$~Hz or $14$~ppb.}

\changed{In FIG.~\ref{fig:spin2}(c), the magnetic field of the permanent magnet assembly was probed by measuring the NMR spectrum with the error in this measurement arising from the uncertainty in determining its centre frequency.} The NMR linewidth is\changed{, in this measurement, is} $\sim 307$~Hz, in contrast to the $\sim 4.25$~kHz linewidth presented in FIG.~\ref{fig:spin1}(a) which was measured \changed{using} a higher RF power. In both NMR spectra, the linewidth is power broadened. The fitting error - specifically the length of the $95\%$ confidence interval - ranges from $13.5$~Hz to $28.8$~Hz, corresponding to an accuracy between $0.80$~ppm and $1.7$~ppm at $\sim 16.961$~MHz.

\changed{To measure the magnetic field drift of the superconducting solenoid, a Ramsey interferometry scheme was performed on a $^{31}$P$^0$ electron spin. Here, instead of using a rotation about $x$ followed by a fixed $\tau$ and subsequent rotation about $y$, both rotations are about $x$. The spin flip probability is then}

\begin{equation}
    P_\mathrm{flip} = \frac{1}{2}+\frac{1}{2}e^{-\chi(\tau)}\cos{\Delta\tau}.
    \label{eq:mag_trace}
\end{equation}

\changed{By varying $\tau$ and measuring $P_\mathrm{flip}$, the resulting data may be fit to Eq.~\ref{eq:mag_trace} and $|\Delta|$ can be determined. To determine the sign of $\Delta$, two different values of $\omega_0$ may be used. The advantage of such a measurement is that it is not limited to small drifts as the scheme used to produce FIG.~\ref{fig:spin2}(a) is. The $95\%$ confidence interval in this scheme was $\sim 10$~kHz at $\sim 43.332$~GHz, corresponding to an accuracy of $0.23$~ppm. This measurement was also applied when measuring the response of the NMR frequency to temperature changes [see FIG.~\ref{fig:spin2}(d)]. In this measurement the $95\%$ confidence interval - ranges from $3.35$~Hz to $43.5$~Hz, corresponding to an accuracy between $0.2$~ppm and $2.5$~ppm at $\sim 16.961$~MHz}

\bibliography{references}

\end{document}